\documentclass[referee]{raa}

\usepackage{amsfonts}
\usepackage{amssymb}
\usepackage{booktabs}
\usepackage{commath}
\usepackage{mathrsfs}
\usepackage{makecell}
\usepackage{threeparttable}
\usepackage{tensor}
\usepackage{longtable}
\usepackage{siunitx}
\usepackage{graphicx,times}
\usepackage{natbib}
\bibpunct{(}{)}{;}{a}{}{,}

\usepackage[graphicx]{realboxes}

\usepackage[colorlinks,linkcolor=blue,anchorcolor=blue,citecolor=blue,urlcolor=blue]{hyperref}

\renewcommand{\tablename}{Table}
\renewcommand{\figurename}{Figure}
\renewcommand{\abstractname}{Abstract}

\begin{document}

   \title{An updated constraint on the Effective Field Theory of Dark Energy}

 \volnopage{ {\bf 20XX} Vol.\ {\bf X} No. {\bf XX}, 000--000}
   \setcounter{page}{1}

   \author{Chi Zheng \inst{1,2}
           \and Wei Liu \inst{1,2} \thanks{Corresponding Author, E-mail: lw980228@mail.ustc.edu.cn}
           \and Zonghao Zhan \inst{1,2}
           \and Wenjuan Fang\inst{1,2} \thanks{Corresponding Author, E-mail: wjfang@ustc.edu.cn}
   }

   \institute{Department of Astronomy, The University of Science
and Technology of China, Hefei 230026, China\\
        \and
        School of Astronomy and Space Sciences, University of Science and Technology of China, Hefei, Anhui, 230026, P.R.China\\
\vs \no
}

\abstract{The Effective Field Theory of Dark Energy (EFTofDE) provides a systematic and model-independent framework to study dark energy (DE) and modified gravity (MG) with one additional scalar degree of freedom. It can describe the known models such as Quintessence, k-essence, DGP, $f(R)$, and Horndeski theories. In this work, we update constraints on EFTofDE by utilizing the most up-to-date public data including the BAO (DESI DR2), CMB (Planck 2018 \& ACT DR6), SNIa (DESY5), weak lensing (DESY3) and full-shape galaxy power (BOSS DR12). We find with the $\Lambda$CDM background, general relativity (GR) is favored by the data, while with the $w0wa$CDM background, slight modification to GR is favored, but still consistent with GR within $1.5\sigma$. We also find the significance level for dynamical DE is greatly reduced within EFTofDE compared to within GR, indicating the degeneracy between dynamical DE and MG.
}

   \authorrunning{C. Zheng, W. Liu, Z.-H. Zhan, \& W.-J.Fang}            
   \titlerunning{An updated constraint on the Effective Field Theory of Dark Energy}  
   \maketitle

%

\section{Introduction}           
\label{sect:Intro}

General relativity (GR), our standard theory of gravity, has passed a wealth of high-precision laboratory and astrophysical tests \citep{Will14,Abbott+16,Will18}. Over the last three decades, the rapid development in observational cosmology, especially the revolutionary discovery of cosmic acceleration in 1998 \citep{Reiss+98,Perlmutter+99}, has extended the test of GR to even larger scales, i.e., the cosmological scales. In the concordance $\Lambda$CDM model, cosmic acceleration is accounted for by the cosmological constant $\Lambda$. However, its 120 orders of magnitude discrepancy with the theoretically predicted value \citep{Weinberg89} makes the exploration of other reasons for cosmic acceleration an active research area \citep{Fri+08,CalKam09}. One of the major effort among these attempts are modification to GR \citep[for reviews on this topic, see][]{Cli+12,Joy+15}, since we have not previously tested GR on cosmological scales.


Cosmological tests of gravity is usually conducted in two different approaches, one is a model-independent phenomenologically parameterized approach \citep[see e.g.][]{Linder05,HuSaw07_new,Zhao+09}, the other is on a model-by-model basis for individual specific models \citep[see e.g.][]{Fang+08,Lombriser+09,Aviles25}. For the former, the effective field theory of dark energy (EFTofDE) \citep{Gubitosi_EFTofDE,Bloomfield+13,EFTofDE_Review} parameterizes deviation from $\Lambda$CDM with one additional scalar degree of freedom, either modification to GR or dark energy, by several time-dependent free functions 
in the Lagrangian, therefore providing a unified description for modified gravity (MG) and dark energy(DE). It can be shown that popular DE or MG models such as the quintessence \citep{Zlatev99}, k-essence \citep{kessence}, DGP \citep{DGP}, $f(R)$ \citep{HuSaw07}, Horndeski \citep{Horndeski74, Deffayet+11} theories can be matched by appropriately setting the free functions. Numerical tools such as \texttt{EFTCAMB} \citep{EFTCAMB}, \texttt{hi\_class} \citep{hiclass} have been developed to solve for the Einstein-Boltzmann equations and cosmic microwave background (CMB) anisotropies in the EFTofDE framework, which facilitates its comparison with observations.

Due to the large number of free functions present in the Lagrangian of the EFTofDE, observational constraints on the framework are usually obtained by making specific assumptions for these functions, and in the literature, constraints have been derived for a variety of different assumptions \citep{Raveri+14,planck2015,Huang16,Frusciante+19,planck2018,DESI_2024FSMG,SDSS_FS_MG} \citep[for a review, see][]{EFTofDE_Review}. Specifically, the Planck team adopted a power-law form for the running of effective Planck mass function $\alpha_M$ in a $\Lambda$CDM background, and obtained constraints consistent with GR both in their 2015 and 2018 results \citep{planck2015,planck2018}.

In this work, we plan to update Planck's constraints by utilizing the most up-to-date public data, especially the recent results of Baryon Acoustic Oscillations (BAO) by the DESI collaboration \citep{DESI_DR2_CosmoConstrain} and weak-lensing measurements by the DES collaboration \citep{DESY3}. The full-shape galaxy power spectrum in redshift space measured by the BOSS collaboration \citep{SDSS_FS_2015} is also included to tighten the constraints on MG \footnote{The DR1 full-shape galaxy power spectrum measured by the DESI collaboration is not publicly available yet when we prepare this work.}. In addition to the $\Lambda$CDM background, we also try to constrain EFTofDE assuming a $w_0w_a$CDM background with a dynamical dark energy whose equation of state is described by the Chevallier-Polarski-Linder (CPL) parameterization \citep{CHEVALLIER_2001,Linder_2003}. Recent analysis by the DESI collaboration \citep{DESI_DR1_BAOCOS, DESI_DR2_CosmoConstrain} has found evidence for such a dynamical dark energy, and it will be interesting to investigate whether the evidence persists when MG is present at the same time.



This paper is organized as follows. In Section \ref{sect:Theory}, we describe the EFTofDE framework for modified gravity. Section \ref{sect:Datasets} summarizes the observational data sets employed in our analysis. In Section \ref{sect:Methodology}, we detail our methodology, including the parameterization choices and inference techniques. Section \ref{sect:Result} presents the constraints on EFTofDE and cosmological parameters, based on various data combinations, compares our findings with previous results, and discusses the implications of our results. In Section \ref{sect:Discussion}, we discuss the limitations and provide directions for future research. Finally, we summarize in Section \ref{sect:Summary}.

\section{Theoretical Framework: Effective Field Theory of Dark Energy}
\label{sect:Theory}

\subsection{Overview of Effective Field Theory of Dark Energy}

EFTofDE \citep{EFTofDE_Review} provides a unified, model-independent framework to describe deviations from General Relativity and $\Lambda$CDM. By focusing on symmetries and relevant degrees of freedom, it encompasses a wide class of dark energy and modified gravity theories—including scalar-tensor models—within a common language characterized by a finite set of parameters directly testable with cosmological observations.

This formalism is particularly well-suited for interpreting data from large-scale structure and CMB experiments, offering a systematic approach to constrain the physics of cosmic acceleration across a broad range of models.

\subsection{Mathematical Formalism}

EFTofDE constructs the most general action consistent with the symmetries of a homogeneous and isotropic background, organized as a derivative expansion in unitary gauge. Based on time diffeomorphism breaking, the formalism directly applies EFT techniques to cosmological perturbations \citep{Gubitosi_EFTofDE,EFTofDE_Review}, with deviations from $\Lambda$CDM parametrized by a set of time-dependent functions that are tightly linked to observables.

The action of the EFTofDE in the unitary gauge is the following \citep{EFTofDE_Review}: 
\begin{equation}
  \begin{aligned}
S_{\text{DE}} &= \int \dif^4 x \sqrt{-g} \bigg[ m_0^2 [1+\Omega(\tau)] \frac{R}{2} + \Lambda(\tau) - a^2 c(\tau) g^{00} \\[0.2cm]
&+\frac{1}{2} M_{2}^{4}(\tau) (a^2\delta g^{00})^2 - \frac{1}{2} \bar{M_{1}}^3(\tau) a^2 \delta g^{00} \delta K - \frac{1}{2} \bar{M_{2}}^2(\tau) (\delta K)^2  \\[0.2cm]
&- \frac{1}{2} \bar{M_3}^2(\tau) \delta \tensor{K}{_\nu^\mu} \delta \tensor{K}{_\mu^\nu}  + \frac{1}{2} \hat{M}^2(\tau) a^2 \delta g^{00} \delta R^{(3)}\\[0.2cm]
&+m_2(\tau) (g^{\mu\nu}+n^{\mu}n^{\nu}) \partial_i (a^2g^{00})\partial^i (a^2g^{00}) + ... \bigg]+ S_{\text{m}}(g_{\mu\nu}, \Psi_{\text{m}}), 
  \label{eq:EFTofDEAction}
  \end{aligned}
\end{equation} 
where $m_0$ is the Planck mass, $R$ is the Ricci scalar, $\delta R^{(3)}$ is the perturbation of the spatial component of the Ricci scalar, $\delta g^{00}$ is defined as $g^{00} + 1, \delta K_{\mu}^{ \nu}$ is the perturbation of the extrinsic curvature, $\delta K$ is its trace, and $S_m(g_{\mu\nu},\Psi_m) $ is the action of matter field except dark energy. Here we can also see nine  functions of the conformal time ($\tau$) in the action modeling the dark energy $\{\Omega(\tau), \Lambda(\tau), c(\tau), M_2(\tau), \bar{M}_1(\tau), \bar{M}_2(\tau), \bar{M}_3(\tau), \hat{M}(\tau), m_2(\tau)\} $. The functions $\{\Lambda (\tau), c(\tau)\}$ affect the background evolution. After specifying the expansion history, these two functions are determined from the Friedman equations. The rest of free functions only change the perturbation evolution. In this work, we study the EFT of dark energy models using \texttt{EFTCAMB} package. 

In the commonly used notation, which is also adopted by the \texttt{EFTCAMB} package, we can rewrite the 6 second-order time dependent $M$ functions in the dimensionless form: 
\begin{equation}
    \begin{aligned}
    \gamma_1 &= \frac{M_{2}^4}{m^2_0 H^2_0}, &
    \gamma_2 &= \frac{\bar{M}_{1}^{3}}{m^2_0 H_0}, &
    \gamma_3 &= \frac{\bar{M}_{2}^{2}}{m^2_0}, \\
    \gamma_4 &= \frac{\bar{M}_{3}^2}{m^2_0}, &
    \gamma_5 &= \frac{\hat{M}^{2}}{m^2_0}, &
    \gamma_6 &= \frac{m_{2}^{2}}{m^2_0}.
    \end{aligned}
    \label{eq:EFTofDEgamma}
\end{equation}
This is also called the EFT-basis, compared to the $\alpha$-basis mentioned in the next section. 

Given an expansion history, and a EFT function $\Omega(\tau)$, we can compute the two functions $\Lambda(\tau)$ and $c(\tau)$. Assuming a flat FLRW metric, we have the Friedmann equations by varying the background part of the action of EFTofDE with respect to the metric \citep{EFTCAMB, EFTCAMBNumericalNotes}: 
\begin{equation}
    \begin{aligned}
    \mathcal{H}^2=\frac{1}{3 m_0^2 (1+\Omega)}(\rho_m+2 c-\Lambda)-\mathcal{H} \frac{\Omega^{\prime}}{(1+\Omega)}, \\
    \mathcal{H}^{\prime}=-\frac{a^2}{6 m_0^2 (1+\Omega)} (\rho_m +3 P_m)-\frac{a^2 (c+\Lambda)}{3 m_0^2 (1+\Omega)}-\frac{\Omega^{\prime\prime}}{2 (1+\Omega)}. 
    \end{aligned}
    \label{eq:EFTofDEHs}
\end{equation} 
Here $\mathcal{H}$ is the conformal Hubble parameter and prime denotes derivative w.r.t.
the conformal time. $\rho_m$ and $P_m$ are the energy density and pressure of the matter components, for which we assume a perfect fluid form with the standard continuity equation: 
\begin{equation}
    \rho_m^{\prime}=-3\mathcal{H} (\rho_m + P_m).
\end{equation}

In the EFTofDE framework, one can adopt a designer approach by fixing the background expansion history a priori—typically to that of a $\Lambda$CDM or $w_0w_a$CDM model, which captures a broad class of dark energy behaviors relevant for full-shape galaxy clustering and weak lensing observations. The background evolution is encoded in three time-dependent functions ${\Omega, \Lambda, c}$, which are related through the following equations: 
\begin{equation}
    \begin{aligned}
    c=-\frac{m_0^2\Omega^{\prime\prime}}{2a^2}+ \frac{m_0^2\mathcal{H}\Omega^{\prime}}{a^2}+ \frac{m_0^2(1+\Omega)}{a^2}(\mathcal{H}^2-\mathcal{H}^{\prime})-\frac{1}{2}(\rho_m + P_m), \\
    \Lambda =-\frac{m_0^2\Omega^{\prime\prime}}{a^2} -\frac{m_0^2\mathcal{H}\Omega^{\prime}}{a^2}- \frac{m_0^2(1+\Omega)}{a^2}(\mathcal{H}^2+2\mathcal{H}^{\prime}) - P_m. 
    \end{aligned}
    \label{eq:EFTofDcandLambda}
\end{equation}
It is convenient to solve for $c$ and $\Lambda$ in terms of $\Omega$, allowing a direct mapping from the specified expansion history to the EFTofDE functions. In the next section, we will see the $M$ functions can be rewriten as the $\alpha$ functions, which have clearer physical meaning.

\subsection{Alternative Parameterisation and Comparison}

Horndeski theory is the most general four-dimensional scalar-tensor theory, characterized by a single scalar field $\phi$ and yielding second-order equations of motion, thus avoiding Ostrogradsky instabilities \citep{PhenoHorndeski}. It includes a broad class of models such as quintessence, $k$-essence, and $f(R)$ gravity. In the EFT framework, it can be formulated in unitary gauge by setting $\phi = tm_0^2$, breaking time diffeomorphism invariance. The EFTofDE action then contains operators invariant under spatial diffeomorphisms, including $g^{00}$, the extrinsic curvature $K_{\mu\nu}$, and the spatial Ricci tensor $R_{\mu\nu}^{(3)}$. After imposing the background equations, the theory can be redefined as four $\alpha$-functions governing cosmological evolution \citep{LSSMG,EFTCAMBNumericalNotes,planck2015}: 
\begin{equation}
    \begin{aligned}
    \alpha_M&=\frac{(M_\ast^2)^{\prime}}{\mathcal{H}M_\ast^2},\\
    \alpha_B&=\frac{m_0^2\Omega^{\prime}+a\bar{M_{1}}^3}{2\mathcal{H}M_\ast^2},\\
    \alpha_K&=\frac{2ca^2+4 a^2M_{2}^4}{\mathcal{H}^2M_\ast^2},\\ 
    \alpha_T&=-\frac{\bar{M}_2^2}{M_\ast^2},
    \end{aligned}
    \label{eq:Horndeskialphas}
\end{equation}
where the effective Planck mass $M_\ast$ is defined as $M_\ast^2=m_0^2(1+ \Omega)+\bar{M}_{2}^{2}$. Note that the factor of $+\frac{1}{2}$ in $\alpha_B$ may vary across different conventions in the literature. Unlike the $M$-functions in the EFT basis which are directly obtained from the EFTofDE action in Equation \ref{eq:EFTofDEAction}, the $\alpha$-functions have a more physical interpretation. For this reason, we adopt both sets of functions to leverage their respective advantages. 

Each $\alpha$ parameter has a distinct physical interpretation \citep{LSSMG}:

\begin{itemize}
    \item[$\bullet$] $\alpha_T$ modifies the speed of gravitational waves; $\alpha_T = 0$ implies propagation at the speed of light.
    \item[$\bullet$] $\alpha_M$ governs the time evolution of the effective Planck mass and impacts gravitational lensing.
    \item[$\bullet$] $\alpha_B$ captures the mixing of kinetic terms between the scalar and metric, affecting structure growth.
    \item[$\bullet$] $\alpha_K$ characterizes the kinetic energy of scalar perturbations and impacts sound speed. 
\end{itemize}

Various models can be recovered by specifying the $\alpha$ parameters. For example \citep{LSSMG,k_Inflation,Essentials_k_essence}: 

\begin{itemize}
    \item[$\bullet$] Quintessence: $\alpha_K = (1 - \Omega_m)(1 + w_X)$, $\alpha_B = \alpha_M = \alpha_T = 0$.
    \item[$\bullet$] $k$-essence: $\alpha_K = \frac{1}{c_s^2}(1 - \Omega_m)(1 + w_X)$, $\alpha_B = \alpha_M = \alpha_T = 0$.
\end{itemize}

\section{Datasets and Observational Probes}
\label{sect:Datasets}

We use a range of cosmological datasets to constrain our models. We incorporate BAO measurements from DESI DR2, which expands sky coverage with over 11,800 tiles ($\sim70\%$ of the full dark-time survey) and improves redshift precision through repeated QSO observations and enhanced BAO modeling via quadrupole terms. For full-shape analysis (FS), we include BOSS DR12 BOSS LOWZ and CMASS samples \citep{SDSS_FS_2015}, using the monopole and quadrupole of the galaxy power spectrum to constrain the growth of structure and the expansion history. Similar as in DESI full-shape studies, we adopt a conservative wavenumber range to exclude nonlinear and systematic-dominated scales. In the following analysis, we consistently combine BAO and FS data to leverage their complementary strengths. BAO measurements provide precise geometric constraints, anchoring dark energy parameters ($w_0$, $w_a$), while FS data, via redshift-space distortions and broadband modeling, probe structure growth, constraining EFTofDE parameters to test deviations from GR. Consistent with prior studies \citep{lu2025preferenceevolvingdarkenergy, SDSS_FS_MG}, this joint analysis mitigates parameter degeneracies, enhancing cosmological constraints without isolated dataset analyses. 

Besides the BAO and FS dataset, we also use several other cosmological datasets, including the cosmic microwave background, Type Ia supernovae, and weak gravitational lensing. We will give a brief description of these datasets in the following sub-sections.

\subsection{Baryon Acoustic Oscillation Data}

Baryon Acoustic Oscillations provide a standard ruler for geometrical measurements, based on the imprint of primordial sound waves in the large-scale distribution of galaxies, which can be used to constrain the Universe's expansion history. The Dark Energy Spectroscopic Instrument (DESI) surveys the large-scale structure of the Universe by collecting spectra of millions of galaxies and quasars. Its first two data releases—DR1 \citep{DESI_DR1} and DR2 \citep{DESI_DR2_I}—offer progressively deeper and broader samples for precision cosmology.

DESI DR2 incorporates data from the first three years, covering 6,671 dark-time and 5,171 bright-time tiles—approximately 2.4- and 2.3-fold increase compared to DR1 respectively—and represents $\sim70\%$ of the full dark-time survey. Besides the increase in the number of galaxies, the DESI DR2 analysis \citep{DESI_DR2_CosmoConstrain} also improves BAO measurements by updating the pipeline for the large-scale structure catalog construction and parameter inference, as well as modifying the theoretical model for BAO. DR2 currently provides the most extensive spectroscopic dataset for BAO measurements. 

In this work, we employ the DESI DR2 BAO likelihood from \texttt{cobaya}, and will denote DESI DR2 BAO data as BAO. 

\subsection{Full-shape Data}

The full-shape of the power spectrum captures the detailed distribution of density fluctuations over a range of scales, containing both geometric information from BAO and dynamical information from Redshift-Space Distortions (RSD). The broadband shape of the power spectrum also provides information about the transfer function, which can also be utilised to constrain cosmological parameters. In this work, we use the monopole and quadrupole moments of the galaxy power spectrum, measured with respect to the line of sight (LOS) \citep{SDSS_FS_2015}, to constrain the expansion history of the Universe, the growth rate of structure, and the transfer function.

While the DESI collaboration's recent full-shape modified gravity analysis \citep{DESI_2024FSMG} used their first-year full-shape dataset to derive cosmological constraints, these data were not yet publicly available during our work; we therefore adopt the BOSS DR12 full-shape measurements \citep{SDSS_FS_2015} as an alternative, leveraging its galaxy power spectrum monopole and quadrupole moments, which provide robust cosmological information—particularly the structure growth rate, $f\sigma_8(z)$. The BOSS DR12 data combine the LOWZ ($z_\mathrm{eff} = 0.32$, including 361,762 galaxies) and CMASS ($z_\mathrm{eff} = 0.57$, including 777,202 galaxies) samples, enabling precise tests of modified gravity scenarios.

In this work, we write the likelihood based on \texttt{desilike} \citep{desilike}, and follow a similar approach to the previous works like \cite{SDSS_FS_MG} and \cite{DESI_2024FSMG}, adopting a conservative wavenumber cut of $k[h \mathrm{Mpc}^{-1}] < 0.20$  to exclude scales affected by systematics or nonlinear modeling uncertainties \citep{DESI_2024FullShape}. This upper bound avoids small-scale nonlinearities where theoretical modeling becomes unreliable. DESI studies have shown that this choice of $k_{\max}$ does not significantly impact parameter constraints, neither the peak nor the width of the posterior \citep{DESI_2024FullModelingParamComp,DESI_2024FullModelingPyBird,DESI_2024FullModelingFOLPS,DESI_2024AnaParamCompFullModelingVelocileptors,DESI_2024EFTComparison}. 

We use the \texttt{velocileptors} code for likelihood evaluation\footnote{DESI collaboration has tested four perturbation theory approaches—\texttt{velocileptors}, \texttt{Folps$\nu$}, \texttt{PyBird}, and \texttt{EFT-GSM}—and found good agreement among them \citep{DESI_2024FullModelingParamComp,DESI_2024FullModelingPyBird,DESI_2024FullModelingFOLPS,DESI_2024AnaParamCompFullModelingVelocileptors,DESI_2024EFTComparison}.}, modeling the redshift-space galaxy power spectrum at one-loop order using Eulerian perturbation theory. For the choice of galaxy bias, stochastic, and counter terms, we adopt the same settings as DESI \citep{DESI_2024AnaParamCompFullModelingVelocileptors}.  Throughout, we refer to the BOSS DR12 full-shape dataset as FS.

\subsection{Cosmic Microwave Background}

The Cosmic Microwave Background (CMB) is a cornerstone in modern cosmology, providing a snapshot of the early universe. It represents the residual thermal radiation from the Big Bang, observed today as a nearly isotropic background at microwave frequencies. 

In this work, we utilize the official Planck 2018 high-$\ell$ TTTEEE (\texttt{plik}) likelihood, complemented by the low-$\ell$ TT (\texttt{Commander}) and EE (\texttt{SimAll}) likelihoods. This is consistent with the choice of the Planck papers \citep{planck2015,planck2018} and DESI full-shape modified gravity paper \citep{DESI_2024FSMG}. In addition to the temperature and polarization anisotropy data, we incorporate measurements of the lensing potential auto-spectrum $C_{\ell}^{\phi\phi}$ from Planck's \texttt{NPIPE} PR4 CMB maps \citep{Carron_2022}, combined with lensing measurements from the Atacama Cosmology Telescope (ACT) Data Release 6 (DR6) \citep{Madhavacheril_2024,Qu_2024}. Specifically, we employ the public implementation in \texttt{cobaya} \citep{actdr6like}, using the \texttt{actplanck\_baseline} option. 

In our analysis, the combination of Planck and ACT DR6 data is denoted as CMB. A version of this dataset without the CMB lensing data is referred to as CMB-nL.

\subsection{Type Ia Supernovae}

Type Ia supernovae, resulting from the thermonuclear explosion of a carbon-oxygen white dwarf in a binary system—via either mass accretion or white dwarf merger \citep{TyIaExpModels}—are pivotal to observational cosmology due to their consistent peak luminosities. Their uniform light curves and spectra make them excellent standard candles for measuring cosmic distances \citep{Riess_1998}. By calibrating peak luminosity through empirical relations such as the Phillips relation \citep{Phillips_1999}, which links luminosity to light curve shape, the intrinsic scatter in observed magnitudes is reduced, enabling precise measurements of the universe's expansion history. This methodology was crucial to the discovery of cosmic acceleration and the inference of dark energy \citep{Perlmutter_1999, Riess_1998}. 

In this work, we employ the DES-SN5YR dataset \citep{DESY5}. It contains a compilation of 194 low-redshift SNIa ($0.025<z<0.1$) and 1635 photometrically classified SNIa covering the range $0.1<z<1.3$. We use the corresponding likelihood in \texttt{cobaya}, which will be denoted as SNIa in the following context.

\subsection{Weak Gravitational Lensing}

Weak gravitational lensing is a powerful probe of the large-scale structure, sensitive to both dark and luminous matter through the deflection of light from distant galaxies by intervening gravitational potentials. It induces small, coherent distortions in galaxy shapes—known as cosmic shear—which can be statistically extracted from large imaging surveys.

We follow the modified gravity analysis of DESY3 \citep{DESY3}, which combines cosmic shear, galaxy-galaxy lensing, and galaxy clustering (the "3×2-point" analysis). To accurately model large-scale galaxy clustering, we avoid the Limber approximation and instead adopt the exact treatment described in \citet{Fang_2020}. The DESY3 3×2-pt data include source galaxies in four redshift bins $[0, 0.36, 0.63, 0.87, 2.0]$ and lens galaxies from the MagLim sample in the first four bins of $[0.20, 0.40, 0.55, 0.70, 0.85, 0.95, 1.05]$.

The weak lensing likelihood is modified from the DES Y1 implementation in \texttt{cobaya} to incorporate DES Y3 data and MG modeling. This setup allows a direct comparison with the DESI full-shape MG analysis \citep{DESI_2024FSMG}. To remain within the linear regime and improve constraints on MG parameters, we apply conservative scale cuts consistent with the DES analysis and set \texttt{use\_Weyl=True}, enabling the use of the Weyl potential power spectrum. We denote the DES Y3 weak lensing dataset as WL throughout.

\section{Methodology}
\label{sect:Methodology}

\subsection{Parameter Estimation Framework}

Bayesian inference updates prior beliefs about model parameters $\theta$ using observational data via Bayes' theorem. The posterior distribution combines the likelihood $P(\text{data}|\theta)$ with prior assumptions; in this work, flat (uninformative) priors are adopted for cosmological parameters.

To sample the posterior, we use Markov Chain Monte Carlo (MCMC) methods, which efficiently explore high-dimensional spaces. Specifically, the Metropolis-Hastings algorithm proposes parameter updates that are accepted with a probability ensuring convergence to the target distribution, enabling robust estimation of parameter uncertainties and correlations.

\subsection{Theoretical Modeling and Computation Tools}

\subsubsection{Einstein-Boltzmann Solvers and EFTofDE}

Einstein-Boltzmann (EB) solvers compute the evolution of cosmological perturbations and observables such as the CMB and large-scale structure. Widely used solvers include \texttt{CAMB} \citep{CAMB,CAMBWeb}, \texttt{CLASS} \citep{class}, and their modified gravity extensions: \texttt{EFTCAMB} \citep{EFTCAMB,EFTCAMB_ConDE,EFTCAMBNumericalNotes}, \texttt{MGCAMB} \citep{MGCAMB}, and \texttt{hi\_class} \citep{hiclass}. 

We use \texttt{EFTCAMB}, an extension of \texttt{CAMB} incorporating the EFTofDE, to model cosmologies beyond General Relativity. It supports a wide range of scalar-tensor and modified gravity models through a modular interface, making it suitable for precision cosmology. One can also use the package \texttt{hi\_class} or its upgraded version \texttt{mochi\_class} for Horndeski theory computation (as shown in the DESI full-shape modified gravity paper \citep{DESI_2024FSMG}), but eventually the two EB-solvers (\texttt{EFTCAMB} and \texttt{hi\_class}) are the same and do not have a clear difference in precision \citep{ComparisonEBSolvers}.

\texttt{EFTCAMB} is highly flexible, allowing seamless switching between models (e.g., Horndeski theory, $f(R)$ Theory) and user-specified parameterizations. Its precision and versatility make it ideal for integration with data analysis tools like \texttt{desilike}. 

\subsubsection{MCMC Sampling with \texttt{cobaya}}

We perform MCMC sampling using \texttt{cobaya}, a modular and parallelizable framework for Bayesian inference. It allows for easy specification of priors, likelihoods, and theories. Our runs use the Metropolis-Hastings sampler from \texttt{CosmoMC}, adapted within \texttt{cobaya}. All MCMC chains under the $\Lambda$CDM background converge with $R-1 < 0.02$, while those under the $w_0w_a$CDM background converge with $R-1 < 0.05$\footnote{We use a higher threshold for $w_0w_a$CDM because it requires significantly longer runtime than $\Lambda$CDM if using the same threshold as $\Lambda$CDM.}.

\subsection{Data Analysis Pipeline}

The analysis is conducted using the modified \texttt{desilike} package \citep{desilike}, which structures the likelihood into three main components: parameterization, theory, and observable. 

\begin{enumerate}
    \item{Parameterization:} Defines the cosmological parameters ($A_s$, $n_s$, $H_0$, $\Omega_b$, $\Omega_{cdm}$, $\tau_{reio}$, $w_0$, $w_a$, $\Omega_0$, $\beta$, see Table \ref{tab:ParametersPrior}) that govern the theoretical predictions. The choice of parameterization is aligned with the scientific goals. 
    \item{Theory and Observable:} We use \texttt{EFTCAMB} to model linear matter perturbations and CMB anisotropies, where  modifications of gravity are incorporated into the Einstein-Boltzmann equations in the EFTofDE approach. The linear matter power spectrum is further input into \texttt{velocileptors} for the modelling of the weakly non-linear galaxy power spectrum in redshift space using the EFTofLSS approach. The linear matter power spectrum is also fed into the theoretical framework of weak lensing used in \citep{DESY3} for the prediction of the $3\times2$ pt statistics, where we incorporate the Weyl potential with non-Limber approximations. We also use \texttt{EFTCAMB} to calculate various kinds of distances at multiple redshifts to compare with those measured from BAO and SNIa. More details on each dataset are provided in their respective subsections in Section \ref{sect:Datasets}.

\end{enumerate}

The pipeline computes model predictions from the template and theory, compares them to observables, and evaluates likelihoods. Parameter estimation is performed via MCMC method. The framework supports model comparison and robustness checks against systematics, enabling iterative refinement. In summary, \texttt{desilike} offers a modular, flexible, and robust pipeline for extracting cosmological constraints from survey data.

\subsection{Parameter Choice And Others}
\label{sec:Horndeski_parameter}
\subsubsection{Parameter Choice on $\alpha$ parameters}
\label{sec:Parameter_choice}

In this work, we adopt the parameterization scheme implemented in \texttt{EFTCAMB} (see Section \ref{sect:Theory} or the \texttt{EFTCAMB} numerical documentation \citep{EFTCAMBNumericalNotes}). It is worth noting that this formalism introduces up to seven EFTofDE free time-dependent functions. Considering the presence of standard cosmological parameters, the effective field theory (EFT) of dark energy still retains significant degrees of freedom. 

To reduce parameter space and computational complexity, we will simplify the EFTofDE model parameterization. We remain in the original class of Horndeski theories, and set $\alpha_T=0$, such that gravitational waves propogage at the speed of light; we turn off high-order EFT operators, which leads to $\alpha_B=-\alpha_M$, then $\alpha_K$ is determined by the functions $\Omega, c$ and $H$ according to Equation (\ref{eq:Horndeskialphas}). As a result, the only additional free function is $\alpha_M$, or equivalently, the EFT basis function $\mathrm{\Omega}$, with the two related through the equation $\alpha_M=\frac{a}{1+\mathrm{\Omega}}\frac{\dif \mathrm{\Omega}}{\dif a}$. In this work, following previous studies \citep{planck2015, planck2018, DESI_2024FSMG}, we adopt a scaling ansatz for $\alpha_M$ with a power-law form: $\alpha_M=\alpha_{M0} a^{\beta}$. Under this setting, we are thus considering a non-minimally coupled "k-essence" type models \citep{planck2015, Sawicki_2013}. Its non-minimal coupling generates second order derivatives in the energy-momentum tensor, reflecting essential characteristics of more general theoretical frameworks \citep{Sawicki_2013}. 

This ansatz allows us to derive two parameterizations for $\mathrm{\Omega}(a)$: an exponential form and a power-law form. The exponential form is obtained by directly integrating the relation between $\alpha_M$ and $\mathrm{\Omega}$, while the power-law form serves as an approximation when $\mathrm{\Omega}_0$ is small. Explicitly, these parameterizations are given by: 
\begin{equation}
    \begin{aligned}
    \mathrm{\Omega}(a)&=\exp{(\mathrm{\Omega}_0 a^{\beta})}-1,\\
    \mathrm{\Omega}(a)&=\mathrm{\Omega}_0 a^{\beta}.
    \end{aligned}
    \label{eq:EFTCAMBParameterization}
\end{equation}
Both two functional forms has been used in previous works \citep{planck2015, planck2018, DESI_2024FSMG}, and in this work, we focus on the exponential form of the function $\Omega(a)$. It is worth noting that, the two forms yield close evolution for $\mathrm{\Omega}(a)$, meaning our choice does not significantly affect the results—a conclusion supported by Figure 8 in \cite{DESI_2024FSMG}, which shows only minor differences across most of the parameter space. 

The \texttt{EFTCAMB} code enforces theoretical and numerical stability conditions to ensure physically viable cosmological solutions. For instance, in the Planck 2015 analysis \citep{planck2015} and the DESI 2024 modified gravity constraints study \citep{DESI_2024FSMG}, the condition $\Omega_0 \geq 0$ was imposed to guarantee compliance with the no-ghost and no-gradient conditions \citep{EFTofDE_Review}. However, this restriction was relaxed in the Planck 2018 analysis \citep{planck2018}. In our work, we focus on physically well-motivated cases in our numerical investigations and provide comparative discussions. Notably, however, in Appendix \ref{sect:Comparison}, where we compare our results with those of Planck 2018, we do not enforce the $\Omega_0 \geq 0$ condition to maintain consistency with their analysis. 

\subsubsection{Choice of Cosmological Parameter Priors}

In Bayesian statistical inference, particularly within Markov Chain Monte Carlo (MCMC) methods, the selection of prior distributions plays a pivotal role in shaping posterior distributions. Priors encapsulate pre-existing knowledge or assumptions about parameters before observational data is incorporated. For cosmological analyses, well-chosen priors are essential not only for precise parameter estimation but also for preserving the physical validity of the results.

\begin{table}
  \centering
  \begin{threeparttable}
    \caption{Prior distributions for cosmological parameters}
    \label{tab:ParametersPrior}
    \begin{tabular}{c|c|c}
      \toprule
      Parameter & Prior Distribution & Notes\\
      \midrule
      $\log(10^{10} A_s)$ & $\mathcal{U}[1.61,3.91]$ & - \\
      $n_s$ & $\mathcal{U}[0.8,1.2]$ & - \\
      $h$ & $\mathcal{U}[0.2,1.0]$ & $H_0/100$ \\
      $\omega_b$ & $\mathcal{U}[0.005,0.1]$ & $\Omega_b h^2$ \\
      $\omega_{cdm}$ & $\mathcal{U}[0.001,0.99]$ & $\Omega_{cdm} h^2$  \\
      $w_0$ & $\mathcal{U}[-3.0,1.0]$ & - \\
      $w_a$ & $\mathcal{U}[-3.0,2.0]$ & - \\
      $\Omega_0$ & $\mathcal{U}[-2.5,7.5]$ & - \\
      $\beta$ & $\mathcal{U}[-1.5,7.5]$ & - \\
      $\tau_{reio}$ & $\mathcal{U}[0.04,0.08]$ & - \\
      \bottomrule
    \end{tabular}
    \begin{tablenotes}[flushleft]
      \item Uniform priors $\mathcal{U}[a, b]$ indicate a flat distribution over the specified interval. 
    \end{tablenotes}
  \end{threeparttable}
\end{table}

To minimize subjective bias, this analysis employs non-informative priors, primarily uniform (flat) distributions for most cosmological parameters. This approach is justified by two key factors: (1) uniform priors maximize entropy in the absence of strong prior knowledge, representing a neutral starting point, and (2) they ensure posterior distributions are predominantly informed by observational data rather than preconceptions. By adopting this systematic strategy, we reduce prior-induced bias, yielding robust and physically meaningful cosmological constraints. We show our choice of parameter priors in Table \ref{tab:ParametersPrior}.

\section{Results}
\label{sect:Result}

Based on the above discussions, our analysis of selected datasets is divided into two parts: one assuming a $\mathrm{\Lambda}$CDM cosmological background and the other assuming a $w_0w_a$CDM background, where we adopt the widely used Chevallier-Polarski-Linder (CPL) parameterization of the dark energy equation of state \citep{CHEVALLIER_2001,Linder_2003}:
\begin{equation}
    \begin{aligned}
    w_{DE}(a)=w_0+w_a(1-a).
    \end{aligned}
    \label{eq:CPLDarkEnergy}
\end{equation}
We adopt the cosmological parameter priors listed in Table \ref{tab:ParametersPrior} and impose the stability condition $\mathrm{\Omega}_0 \geq 0$ to restrict our analysis to physically viable regions. 

We derive cosmological constraints using four dataset combinations: BAO+FS+CMBnL, BAO+FS+CMB+SNIa, BAO+FS+CMBnL+WL and BAO+FS+CMBnL+SNIa+WL. This tiered approach allows us to test the consistency of results across different probes, isolate potential systematics, and assess tensions with both the $\mathrm{\Lambda}$CDM and $w_0w_a$CDM backgrounds. The final combination, incorporating BAO, FS, SNIa, CMBnL and WL, provides the most stringent constraints. 

\subsection{Results with $\mathrm{\Lambda}$CDM Background Expansion}

We consider a modified gravity model within the framework of EFTofDE, evolving on a standard $\Lambda$CDM background. The exponential parameterization form $\Omega_0(a)=\exp{(\Omega_0 a^{\beta})}-1$ we adopted has been described in detail in section \ref{sec:Horndeski_parameter}. By introducing additional scalar degrees of freedom or corrections to the Einstein-Hilbert action, this framework provides a theoretical avenue to explain the late-time accelerated expansion of the Universe. Within this setup, we carry out MCMC computations and perform a likelihood analysis using observational data to constrain key cosmological parameters. The results are summarized in Table \ref{tab:LCDMResult}.

\begin{figure}[h]
    \centering
    \includegraphics[width=0.29\columnwidth]{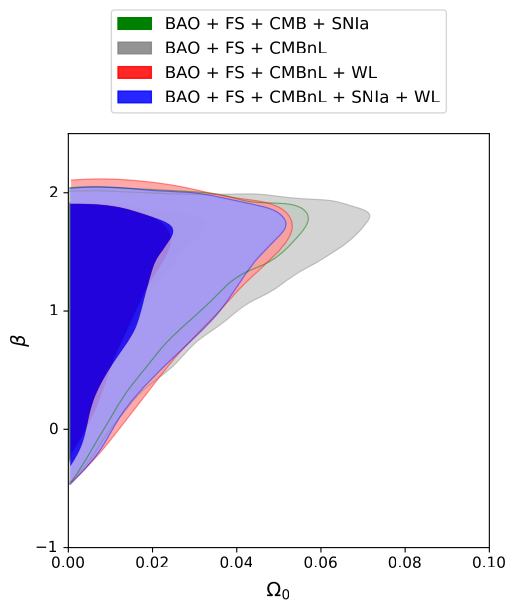}
    \includegraphics[width=0.29\columnwidth]{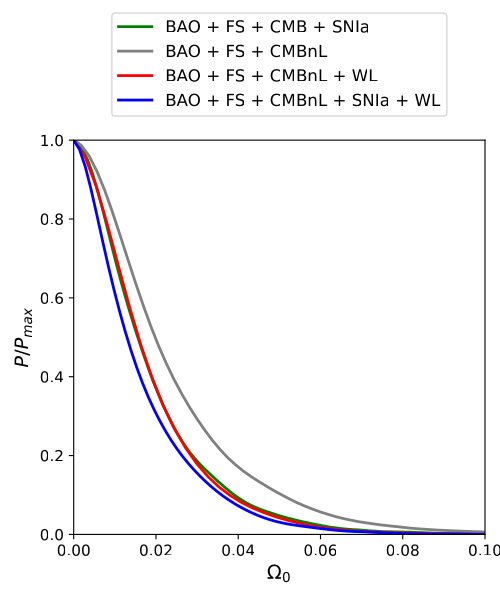}
    \includegraphics[width=0.285\columnwidth]{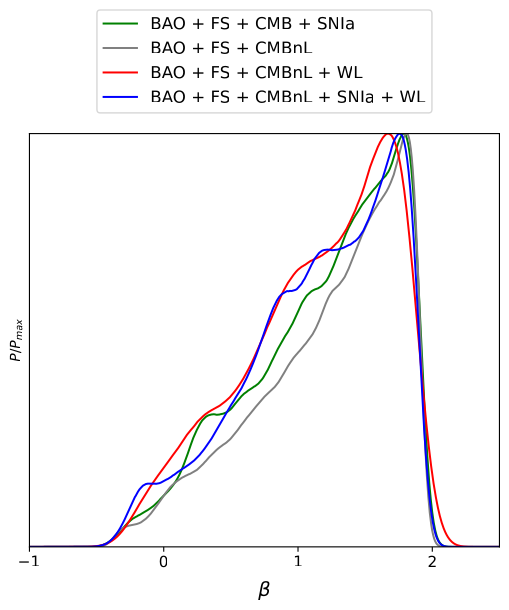}
    \caption{Left: The $68\%$ and $95\%$ contours for EFTofDE parameters $\Omega_0$ and $\beta$, assuming a $\Lambda$CDM background. Center \& Right: The marginalized posterior distribution for the EFT parameter $\Omega_0$ and $\beta$ in $\Lambda$CDM background. To restrict our analysis to physically viable models, we impose the stability condition $\Omega_0 \geq 0$. The results indicate no preference for deviation from standard General Relativity.
    }
    \label{fig:EFT-Fig-Result}
\end{figure}

The combination of our selected data combination imposes tight constraints on the two studied parameters of the EFTofDE model: $\Omega_0$ and $\beta$. Even with just BAO+FS+CMBnL data, the parameter space is already significantly restricted. We find $95\%$ upper bound of $0.056$ for $\Omega_0$ with BAO+FS+CMBnL data combination. The addition of SNIa and CMB lensing data or WL data further improves these constraints, tighten the upper bound to $0.044$ and $0.041$, respectively. Furthermore, the most stringent bounds of $0.038$ is obtained when all datasets are combined. This demonstrates that joint analyses incorporating multiple cosmological probes can effectively reduce parameter uncertainties and enhance the constraining power on the model.

\begin{table}
  \centering
  \begin{threeparttable}
    \caption{Marginalized means and $1\sigma$ uncertainties of cosmological parameters. For the EFTofDE parameter $\Omega_0$, for which the stability condition $\Omega_0 \geq 0$ is enforced, we show its $95\%$ upper bound. Here, we assume a $\mathrm{\Lambda}$CDM background.}
    \label{tab:LCDMResult}
    \begin{tabular}{c|c|c|c|c}
      \toprule
      Parameters & \makecell{BAO+FS+CMB\\+SNIa} & BAO+FS+CMBnL & \makecell{BAO+FS+CMBnL\\+WL} & \makecell{BAO+FS+CMBnL\\+WL+SNIa}\\
      \midrule
      $n_\mathrm{s}$ & $0.9683\pm 0.0034$ & $0.9690\pm 0.0034$ & $0.9688\pm 0.0034$ & $0.9680\pm 0.0032$\\

      $h$ & $0.6849^{+0.0028}_{-0.0031}$ & $0.6869^{+0.0031}_{-0.0027}$ & $0.6855\pm 0.0029$ & $0.6837\pm 0.0026$\\

      $\omega_\mathrm{b}$ & $0.02248\pm 0.00013$ & $0.02249\pm 0.00013$ & $0.02247\pm 0.00013$ & $0.02243\pm 0.00012$\\

      $\omega_\mathrm{c}$ & $0.11863\pm 0.00068$ & $0.11814\pm 0.00067$ & $0.11846\pm 0.00060$ & $0.11884\pm 0.00060$\\

      $\mathrm{\Omega}_0$ [95\% C.L.] & $<0.044$ & $<0.056$ & $<0.041$ & $<0.038$\\

      $\beta$ & $1.19^{+0.72}_{-0.27}$ & $1.22^{+0.69}_{-0.24}$ & $1.16^{+0.72}_{-0.33}$ & $1.16^{+0.73}_{-0.29}$\\

      \bottomrule
    \end{tabular}
  \end{threeparttable}
\end{table}

It is important to note that the EFTofDE parameter $\Omega_0 = 0$ corresponds to the standard General Relativity limit. As shown in Figure \ref{fig:EFT-Fig-Result}, current data favors the value of $\Omega_0$ peaked at zero, indicating no preference for deviation from GR. This suggests that while the EFTofDE framework, by introducing additional degrees of freedom, provides a more flexible framework that can accommodate potential departures from GR, the standard General Relativity remains consistent with observations. The posterior distribution of $\Omega_0$, presented in Figures \ref{fig:EFT-Fig-Result}, peak at zero, but exhibits a skewness hinting at a preference for slightly negative values, which is in accordence with the result when we relax the stability condition (see Appendix \ref{sect:Comparison}). 

For the EFTofDE parameter $\beta$, the posterior distribution remains consistently centered around unity across all dataset combinations. We find a weak positive correlation between $\Omega_0$ and $\beta$ when assuming $\Omega_0 \geq 0$ in Figure \ref{fig:EFT-Fig-Result}, however, the correlation disappears and can even become negative in Figure \ref{fig:Pl-DESI}, when we disregrad of the stability condition $\Omega_0 \geq 0$. Given the function form $\Omega(a) = \exp(\Omega_0 a^\beta) - 1$ and the relatively small value of $\Omega_0$, this indicates that the data favor an approximately linear evolution of $\Omega(a)$ with the scale factor $a$. 

Our results can be compared with those from Planck 2015 \citep{planck2015}, which obtained $\alpha_{M0} < 0.062$ ($95\%$ C.L.)\footnote{We have the definition $\Omega_0 = \frac{\alpha_{M0}}{\beta}$, such that $\Omega(a) = \exp(\Omega_0 a^\beta) - 1=\exp(\frac{\alpha_{M0}}{\beta} a^\beta) - 1$. } for the combined dataset CMB (Planck TT+TE+EE)+BAO (6dFGS+SDSS MGS+BOSS DR11)+SNIa (JLA)+$H_0$ prior(\cite{Riess_2011}). In our analysis, we find a tighter constraint of $\alpha_{M0} < 0.050$ for the data combination BAO+FS+CMB+SNIa, demonstrating that the inclusion of the full-shape dataset along with updated SNIa, BAO and CMB data improves the precision of the EFTofDE parameter constraints. To make a comparison between our result and DESI result, we have derived, based on constraints from BAO (DESI DR1)+FS (DESI DR1)+CMB+SNIa in the DESI full-shape modified gravity analysis \citep{DESI_2024FSMG}, the contraints $\Omega_0 < 0.0494$ \footnote{DESI only reports their constraint $\Omega_0 < 0.0412$ for the power-law parameterization. Here, we derive the DESI constraint for the exponential parameterization based on the constraints provided in their paper: $\alpha_{M0}<0.0445$ in the $95\%$ C.L., and $\beta$ is infered to be $1.22^{+0.71}_{-0.23}$.}. Although our study uses an older FS dataset, the addition of newer BAO measurements (DESI DR2) allows us to achieve constraints with slightly higher precision. This underscores the importance of BAO data in refining the bounds on our parameter. Despite differences in the datasets employed, both analyses lead to consistent conclusions regarding the constraints on EFT parameters, reinforcing the robustness of these results across different observational combinations.

\subsection{Results with $w_0w_a$CDM Background Expansion}

\begin{figure}[h]
    \centering
    \includegraphics[width=8cm]{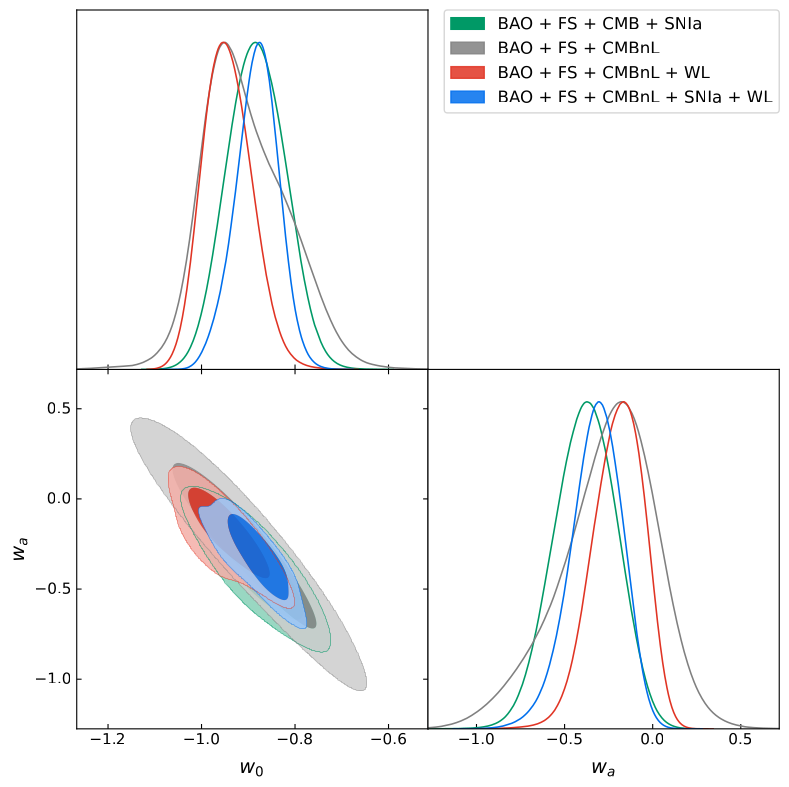}
    \caption{The $68\%$ and $95\%$ contours for dark energy parameters $w_0$ and $w_a$, assuming a $w_0w_a$CDM background. To restrict our analysis to physically viable models, we impose the stability condition $\Omega_0 \geq 0$. The results indicate suggests a more dynamic dark energy.}
    \label{fig:EFTw0wa}
\end{figure}

In this subsection, we study the modified gravity model within EFTofDE, now embedded in a $w_0w_a$CDM background rather than the $\Lambda$CDM background. This alternative background alters the cosmic expansion history and leads to different constraints on cosmological parameters. The resulting parameter estimates are summarized in Table \ref{tab:w0waResult}.

\begin{figure}[h]
    \centering
    \includegraphics[height=0.35\columnwidth]{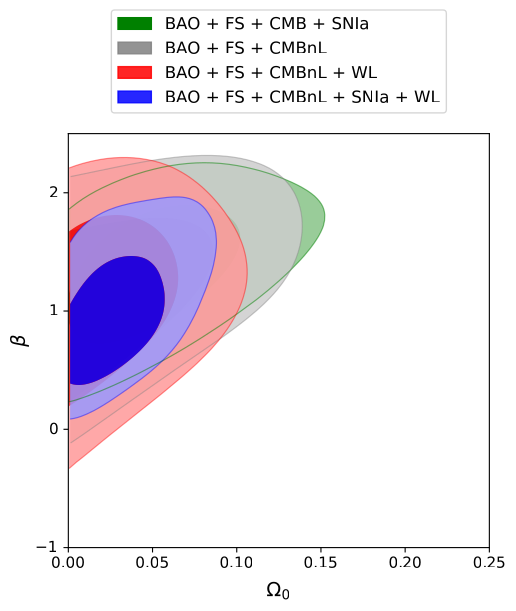}
    \includegraphics[height=0.35\columnwidth]{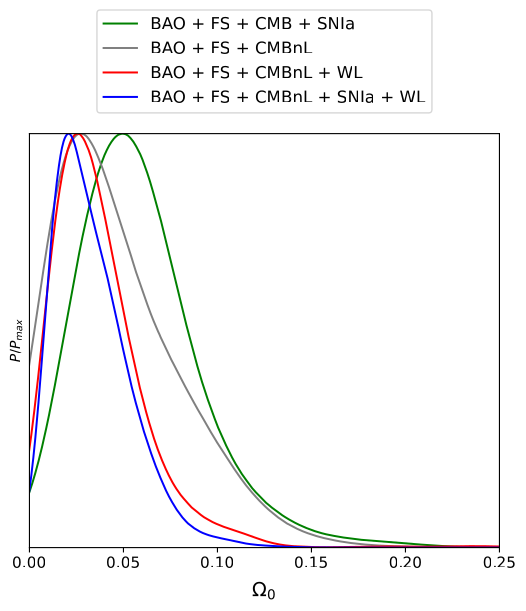}
    \includegraphics[height=0.35\columnwidth]{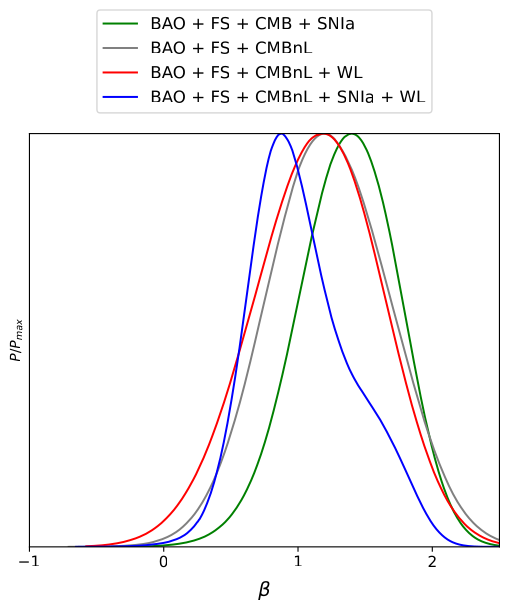}
    \caption{Left: The $68\%$ and $95\%$ contours for EFTofDE parameters $\Omega_0$ and $\beta$, assuming a $w_0w_a$CDM background. Center \& Right: The marginalized posterior distribution for the EFT parameter $\Omega_0$ and $\beta$ in $w_0w_a$CDM background. To restrict our analysis to physically viable models, we impose the stability condition $\Omega_0 \geq 0$. While the result shows a non-negative best-fit value of $\Omega_0$, different from the $\Lambda$CDM background, the results still indicate no statistically significant deviation from standard General Relativity.}
    \label{fig:EFTw0wa-Fig-Result}
\end{figure}

We perform a joint analysis using various combinations of data sets and compare the results with previous outcomes. When comparing the results in Table \ref{tab:w0waResult} with those under the $\Lambda$CDM background in Table \ref{tab:LCDMResult}, we find that the constraints in the $w_0w_a$CDM background are consistently weaker. This is expected, as the evolving dark energy equation of state introduces additional degrees of freedom, reducing the constraining power on the cosmological parameters.  

Table \ref{tab:w0waResult} and Figure \ref{fig:EFTw0wa-Fig-Result} show that the best-fit value of the EFTofDE parameter $\Omega_0$ is slightly positive, deviating from the General Relativity limit of $\Omega_0 = 0$ and thus differing slightly from the results with the $\Lambda$CDM background in the previous sub-section. However, this deviation is not statistically significant, and all data sets remain consistent with GR within $1.5\sigma$. This result is consistent with the DESI full-shape modified gravity analysis \citep{DESI_2024FSMG}, which used the combined dataset BAO(DESI DR1) + FS(DESI DR1) + CMBnL + WL and reported $\Omega_0 = 0.043^{+0.016}_{-0.031}$. The mild preference for a positive $\Omega_0$ may indicate a degeneracy between modified gravity effects and the background evolution of dark energy. While not conclusive, this trend hints at a possible departure from standard GR with the $w_0$–$w_a$ background. Nevertheless, the overall evolution of $\Omega(a)$ remains approximately linear, with best-fit values still consistent with the $\Lambda$CDM background case.

\begin{table}
  \centering
  \begin{threeparttable}
    \caption{Marginalized means and $1\sigma$ uncertainties of cosmological parameters. Here, we assume a $w_0w_a$CDM background.}
    \label{tab:w0waResult}
    \begin{tabular}{c|c|c|c|c}
      \toprule
      Parameters & \makecell{BAO+FS+CMB\\+SNIa} & BAO+FS+CMBnL & \makecell{BAO+FS+CMBnL\\+WL} & \makecell{BAO+FS+CMBnL\\+WL+SNIa}\\
      \midrule
      $n_\mathrm{s}$ & $0.9663\pm 0.0038$ & $0.9684^{+0.0031}_{-0.0050}$ & $0.9668^{+0.0043}_{-0.0037}$ & $0.9680^{+0.0033}_{-0.0029}$\\

      $h$ & $0.6779\pm 0.0086$ & $0.6771\pm 0.0096$ & $0.682^{+0.015}_{-0.0083}$ & $0.6742^{+0.0027}_{-0.0073}$\\

      $\omega_\mathrm{b}$ & $0.02246\pm 0.00012$ & $0.02248\pm 0.00014$ & $0.02242\pm 0.00014$ & $0.02246\pm 0.00014$\\

      $\omega_\mathrm{c}$ & $0.11920\pm 0.00090$ & $0.1185\pm 0.0012$ & $0.11913^{+0.00075}_{-0.00093}$ & $0.11884\pm 0.00083$\\

      $w_0$ & $-0.883\pm 0.059$ & $-0.901^{+0.070}_{-0.11}$ & $-0.945^{+0.048}_{-0.056}$ & $-0.882^{+0.049}_{-0.042}$\\

      $w_a$ & $-0.38\pm 0.17$ & $-0.27^{+0.31}_{-0.20}$ & $-0.19^{+0.16}_{-0.13}$ & $-0.32^{+0.16}_{-0.13}$\\

      $\Omega_{0}$ & $0.057^{+0.024}_{-0.036}$ & $0.044^{+0.024}_{-0.042}$ & $0.036^{+0.015}_{-0.028}   $ & $0.033^{+0.013}_{-0.024}$\\

      $\beta$ & $1.37^{+0.38}_{-0.34}$ & $1.22\pm 0.41$ & $1.14\pm 0.43$ & $1.03^{+0.28}_{-0.46}$\\
      \bottomrule
    \end{tabular}
  \end{threeparttable}
\end{table}

\begin{table}
  \centering
  \begin{threeparttable}
    \caption{Summary of the difference in the effective $\Delta\chi^2$ value, for the best-fit $w_0w_a$CDM background relative to the standard $\mathrm{\Lambda}$CDM background, and the corresponding significance levels. }
    \label{tab:deltachiSquare}
    \begin{tabular}{c|c|c}
      \toprule
      Dataset & $\Delta\chi_{MAP}^2$ & Significance \\
      \midrule
      BAO+FS+CMB+SNIa & $-7.37$ & $2.24\sigma$\\
      BAO+FS+CMBnL & $-2.67$ & $1.12\sigma$\\
      BAO+FS+CMBnL+WL & $-3.53$ & $1.37\sigma$\\
      BAO+FS+CMBnL+SNIa+WL & $-8.42$ & $2.44\sigma$\\
      \bottomrule
    \end{tabular}
  \end{threeparttable}
\end{table}

In addition to the constraints on the EFTofDE parameters, our analysis also yields insights into dark energy dynamics. To quantitatively assess whether an extended dark energy model provides a better fit than $\Lambda$ within the EFTofDE framework, we analyze the difference in the $\chi^2$ values at the maximum a posteriori (MAP) points for these two backgrounds (this method is commonly employed in cosmological studies, like \cite{planck2018} and \cite{DESI_2024FSMG}). Since $\Lambda$CDM is a special case of $w_0w_a$CDM with $(w_0, w_a) = (-1, 0)$, Wilks' theorem \citep{WilksTheorem} implies that—under the null hypothesis—$\Delta\chi^2_{\mathrm{MAP}}$ follows a $\chi^2$ distribution with two degrees of freedom, assuming Gaussian errors and well-estimated uncertainties. For easier interpretation, we convert $\Delta\chi^2_{\mathrm{MAP}}$ into an equivalent Gaussian significance $N\sigma$, defined via: $CDF_{\chi^2}(\Delta\chi^2_{\mathrm{MAP}}|2 \mathrm{dof})=\frac{1}{\sqrt{2 \pi}}\int^{N}_{-N} e^{-t^2/2}\dif t$, where $CDF_{\chi^2}$ is the cumulative distribution of $\chi^2$ for two degrees of freedom. A negative $\Delta\chi^2_{\mathrm{MAP}}$ suggests that the $w_0w_a$CDM model, with its additional free parameters ($w_0$ and $w_a$), achieves a better fit to the data. The magnitude of $\Delta\chi^2_{\mathrm{MAP}}$ can be used to infer the strength of this preference. 

Table \ref{tab:deltachiSquare} presents the values of $\Delta\chi^2_{\mathrm{MAP}}$ and their corresponding significance levels for various dataset combinations. In all cases, $\Delta\chi^2_{\mathrm{MAP}} < 0$, indicating a preference for the $w_0w_a$CDM model over $\Lambda$CDM. The strongest deviation ($2.44\sigma$) arises from the full combination BAO+FS+CMBnL+SNIa+WL, underscoring the enhanced sensitivity to dark energy dynamics when CMB lensing, SNIa, and WL data are included. Removing CMB lensing and SNIa reduces the significance to $1.12\sigma$, while adding WL modestly improves it to $1.37\sigma$, highlighting the individual contributions of each dataset. These trends are consistent with Figure \ref{fig:EFTw0wa}. While combinations such as BAO+FS+CMBnL+WL (BAO+FS+CMBnL) yields only weak ($<2\sigma$) evidence, the inclusion of SNIa (SNIa and CMB lensing) increases the significance to above $>2\sigma$ (consistent with the findings by DESI collaboration using the DES year 5 supernovae). For comparison, the DESI DR2 key paper \citep{DESI_DR2_CosmoConstrain} reports $4.2\sigma$ with BAO+CMB+SNIa with GR, while we obtain $2.2\sigma$ for BAO+FS+CMB+SNIa with EFTofDE. Therefore, we conclude that with MG present which is further constrained by the extra growth information contained in the BOSS full-shape data, the preference for dynamical dark energy is greatly reduced.

Throughout our analysis, we have consistently used the combination of BAO and FS, as generally speaking, these two probes are available from the same spectroscopic galaxy sample, in addition to the fact that they offer complementary geometric and growth constraints. Should we neglect FS, we would expect the contours of $w_0, w_a$ remain largely unchanged, judging from the contours for BAO+CMB+SN1a with and without DESI FS data included from \cite{DESI_2024CCoFullShape}. While with either FS or BAO taken away from the data combinations, we expect the constraints on $\Omega_0$ will become weaker, thus probably reducing the significance level of deviation from GR.

\section{Discussion}
\label{sect:Discussion}

This study presents constraints on EFTofDE models using full-shape galaxy clustering data. However, several limitations and opportunities for further progress remain.

First, we restrict our analysis to the function $\Omega(a)$—one of nine functions introduced in the EFTofDE action—using a simplified parameterization. While this choice ensures numerical stability, it omits other operators, such as those modifying the kinetic term or introducing gravitational slip, which are essential to capture the full range of viable Horndeski models. Extending the EFT parameter space in future work could uncover richer signatures of modified gravity, though it might lead to weaker constraints due to additional degrees of freedom. 

Second, although we successfully integrated \texttt{EFTCAMB} into the \texttt{cosmodesi} pipeline and applied it to BOSS full-shape data, the analysis is limited to a conservative wavenumber range of $k[h\mathrm{Mpc}^{-1}] < 0.20$. This cut, common in large-scale structure studies, avoids nonlinear systematics but excludes small-scale information that could enhance sensitivity to gravity modifications. Increasing $k_{max}$ would improve sensitivity to such effects, but at the cost of potential biases from nonlinearities and modeling uncertainties. While \texttt{velocileptors} provides partial modeling of nonlinearities, more advanced techniques—such as the halo model reaction framework\citep{Bose_2023, Bose_2016, Carlson_2009} or higher-order perturbation theory—may allow for the exploration of smaller-scale information. However, small-scale analyses remain challenging due to the influence of baryonic physics, galaxy formation, and observational systematics. Careful modeling and simulation of these effects are crucial for obtaining robust results. 

Looking forward, upcoming DESI full-shape measurements will play a key role. While we focus here on the well-studied BOSS dataset, DESI's first three-year BAO data have already been released \citep{DESI_DR1,DESI_DR2_I}, and full-shape analyses from the three-year sample are ongoing. These new data will offer significantly improved precision on cosmic large-scale structure, enabling tighter and more comprehensive tests of modified gravity.

In summary, this work lays the foundation for future EFTofDE analyses by combining physically motivated models with high-precision data and robust analysis pipeline. We will broaden the EFTofDE parameter space, improve nonlinear modeling of the power spectrum and apply the updated methodology to the next-generation surveys in future work.

\section{Summary}
\label{sect:Summary}

This work employs the Effective Field Theory of Dark Energy (EFTofDE) to model cosmic acceleration and investigates how high-precision data—particularly from BOSS and DESI—can constrain modified gravity parameters. By applying theoretical models to observables and integrating MCMC analysis techniques, we bridge the gap between fundamental physics and cosmological data.

We perform an analysis using DESI BAO, full-shape BOSS clustering measurements, Planck CMB, DES weak lensing, and DES supernovae—ensures across multiple redshift bins, incorporating full covariance matrices, perturbation theory, and nuisance parameter modeling. This yields tight constraints on both standard cosmological parameters ($H_0$, $\Omega_m$, $\sigma_8$, etc. ) and modified gravity parameters ($\Omega_0$, $\beta$). The MCMC results show no significant deviation from standard General Relativity but mildly favor a dynamical dark energy component, consistent with DESI DR2 BAO results \citep{DESI_DR2_CosmoConstrain}. 

This study highlights EFTofDE as a practical, interpretable framework for testing gravity with large-scale structure. And the pipeline used here can be a starting point for future analyses. Upcoming surveys (e.g., DESI, LSST, Euclid), combined with novel probes such as gravitational wave probes and more advanced analysis methods, will further enhance this framework. Promising directions include multi-probe consistency tests, redshift-resolved EFT parameter evolution, and multi-messenger constraints on gravity etc.

In conclusion, this work uses a flexible and data-driven approach to test dark energy and gravity, advancing toward a precision cosmological model grounded in fundamental theory.

\appendix

\section{Comparison with Planck 2018 Results}
\label{sect:Comparison}
For a better comparison with constraints from Planck 2018, we here follow their analysis choice, by removing the stability conditions such as the no-ghost and no-gradient criteria, allowing $\Omega_0 < 0$ and admitting scenarios with a decreasing Planck mass over time. Although such models may raise theoretical concerns, they enable broader exploration of the parameter space.

\begin{figure}[h]
    \centering
    \includegraphics[width=8cm]{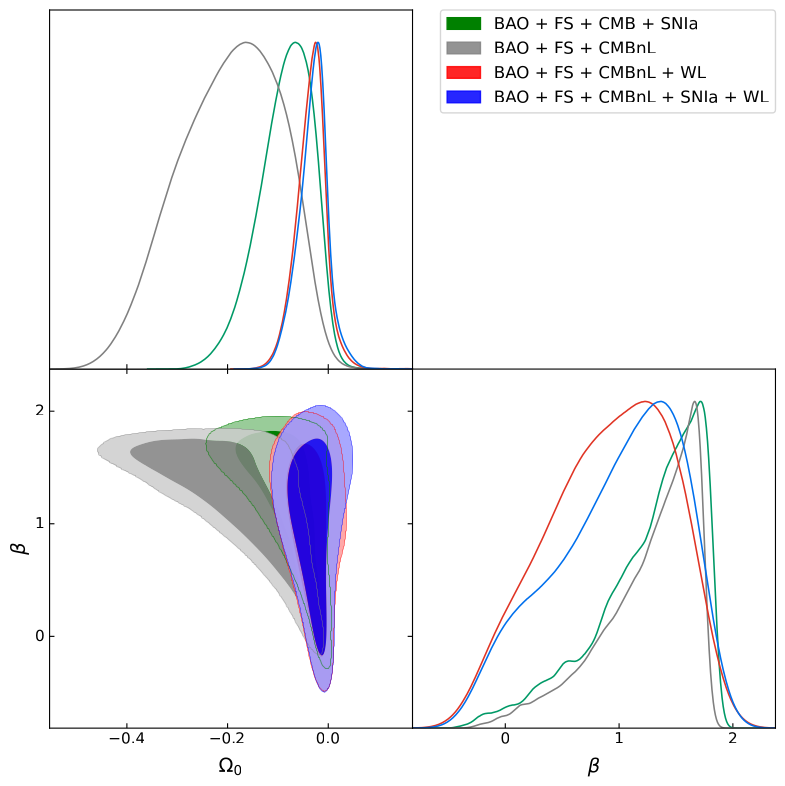}
    \caption{The $68\%$ and $95\%$ contours for EFTofDE parameters $\Omega_0$ and $\beta$, assuming a $\Lambda$CDM background. The stability condition $\Omega_0 \geq 0$ is not imposed, to maintain consistency with the conditions used in the Planck 2018 analysis. }
    \label{fig:Pl-DESI}
\end{figure}

Our constraints on $\Omega_0$ are consistent with those from Planck 2018, indicating that this parameter is robustly constrained across different datasets. Since the EFTofDE framework reduces to GR when $\Omega_0 = 0$, these constraints serve as a direct test of possible deviations from GR.

To quantify the statistical significance of such deviations, we evaluate the preference for $\Omega_0 \neq 0$ under various data combinations. Using BAO+FS+CMB+SNIa, we find a $0.84\sigma$ deviation; this increases to $1.05\sigma$ with BAO+FS+CMBnL, but decreases to $0.53\sigma$ when weak lensing is added (BAO+FS+CMBnL+WL). For the full combination (BAO+FS+CMBnL+SNIa+WL), the deviation is $0.78\sigma$. In comparison, Planck 2018 reported $0.9\sigma$ and $2.1\sigma$ deviations using CMBnL+BAO/RSD+WL and CMBnL only datasets, respectively \citep{planck2018}.

Given these modest deviations, along with the results presented in Section \ref{sect:Result}, we conclude that current data do not provide statistically significant evidence for deviations from GR, whether with or without the stability condition of $\Omega_0 \geq 0$. Further data and analysis will be necessary to robustly test potential departures from GR.

\normalem
\begin{acknowledgements}
We thank Jiaming Pan, Bin Hu, Mustapha Ishak for useful conversations. This work is supported by the National Key R\&D Program of China Grant No. 2022YFF0503404 and No. 2021YFC2203100, by the National Natural Science Foundation of China Grants No. 12173036 and 11773024, by the China Manned Space Program with Grant No. CMS-CSST-2025-A04, by Cyrus Chun Ying Tang Foundations, and by the 111 Project for "Observational and Theoretical Research on Dark Matter and Dark Energy" (B23042).

\end{acknowledgements}

\bibliographystyle{raa}
\bibliography{bibtex}

\begin{thebibliography}{78}
\providecommand\natexlab[1]{#1}
\providecommand\JournalTitle[1]{#1}

\bibitem[{Abbott} {et~al.}(2016)]{Abbott+16}
{Abbott}, B.~P., {Abbott}, R., {Abbott}, T.~D., {et~al.} 2016, \prl, 116, 061102

\bibitem[{Abbott} {et~al.}(2023)]{DESY3}
{Abbott}, T.~M.~C., {Aguena}, M., {Alarcon}, A., {et~al.} 2023, \prd, 107, 083504

\bibitem[{ACT Collaboration}(2023)]{actdr6like}
{ACT Collaboration}. 2023, ACT DR6 Lensing Likelihood, \url{https://github.com/ACTCollaboration/act_dr6_lenslike}

\bibitem[{Adame} {et~al.}(2025{\natexlab{a}})]{DESI_2024FullShape}
{Adame}, A.~G., {Aguilar}, J., {Ahlen}, S., {et~al.} 2025{\natexlab{a}}, {DESI 2024 V: Full-Shape galaxy clustering from galaxies and quasars}

\bibitem[{Adame} {et~al.}(2025{\natexlab{b}})]{DESI_DR1_BAOCOS}
{Adame}, A.~G., {Aguilar}, J., {Ahlen}, S., {et~al.} 2025{\natexlab{b}}, \jcap, 2025, 021

\bibitem[{Adame} {et~al.}(2025{\natexlab{c}})]{DESI_2024CCoFullShape}
---. 2025{\natexlab{c}}, {DESI 2024 VII: cosmological constraints from the full-shape modeling of clustering measurements}

\bibitem[{Armend{\'a}riz-Pic{\'o}n} {et~al.}(1999)]{k_Inflation}
{Armend{\'a}riz-Pic{\'o}n}, C., {Damour}, T., \& {Mukhanov}, V. 1999, Physics Letters B, 458, 209

\bibitem[{Armendariz-Picon} {et~al.}(2000)]{kessence}
{Armendariz-Picon}, C., {Mukhanov}, V., \& {Steinhardt}, P.~J. 2000, \prl, 85, 4438

\bibitem[{Armendariz-Picon} {et~al.}(2001)]{Essentials_k_essence}
{Armendariz-Picon}, C., {Mukhanov}, V., \& {Steinhardt}, P.~J. 2001, \prd, 63, 103510

\bibitem[{Aviles}(2025)]{Aviles25}
{Aviles}, A. 2025, \prd, 111, L021301

\bibitem[{Bellini} \& {Sawicki}(2014)]{LSSMG}
{Bellini}, E., \& {Sawicki}, I. 2014, \jcap, 2014, 050

\bibitem[{Bellini} {et~al.}(2018)]{ComparisonEBSolvers}
{Bellini}, E., {Barreira}, A., {Frusciante}, N., {et~al.} 2018, \prd, 97, 023520

\bibitem[{Bloomfield} {et~al.}(2013)]{Bloomfield+13}
{Bloomfield}, J., {Flanagan}, {\'E}.~{\'E}., {Park}, M., \& {Watson}, S. 2013, \jcap, 2013, 010

\bibitem[{Bose} \& {Koyama}(2016)]{Bose_2016}
{Bose}, B., \& {Koyama}, K. 2016, \jcap, 2016, 032

\bibitem[{Bose} {et~al.}(2023)]{Bose_2023}
{Bose}, B., {Tsedrik}, M., {Kennedy}, J., {et~al.} 2023, \mnras, 519, 4780

\bibitem[{Caldwell} \& {Kamionkowski}(2009)]{CalKam09}
{Caldwell}, R.~R., \& {Kamionkowski}, M. 2009, Annual Review of Nuclear and Particle Science, 59, 397

\bibitem[{Carlson} {et~al.}(2009)]{Carlson_2009}
{Carlson}, J., {White}, M., \& {Padmanabhan}, N. 2009, \prd, 80, 043531

\bibitem[{Carron} {et~al.}(2022)]{Carron_2022}
{Carron}, J., {Mirmelstein}, M., \& {Lewis}, A. 2022, \jcap, 2022, 039

\bibitem[{Chevallier} \& {Polarski}(2001)]{CHEVALLIER_2001}
{Chevallier}, M., \& {Polarski}, D. 2001, International Journal of Modern Physics D, 10, 213

\bibitem[{Clifton} {et~al.}(2012)]{Cli+12}
{Clifton}, T., {Ferreira}, P.~G., {Padilla}, A., \& {Skordis}, C. 2012, \physrep, 513, 1

\bibitem[{de Boe} {et~al.}(2024)]{PhenoHorndeski}
{de Boe}, D., {Ye}, G., {Renzi}, F., {et~al.} 2024, {Phenomenology of Horndeski gravity under positivity bounds}

\bibitem[{Deffayet} {et~al.}(2011)]{Deffayet+11}
{Deffayet}, C., {Gao}, X., {Steer}, D.~A., \& {Zahariade}, G. 2011, \prd, 84, 064039

\bibitem[{DES Collaboration} {et~al.}(2024)]{DESY5}
{DES Collaboration}, {Abbott}, T.~M.~C., {Acevedo}, M., {et~al.} 2024, {The Dark Energy Survey: Cosmology Results with {\ensuremath{\sim}}1500 New High-redshift Type Ia Supernovae Using the Full 5 yr Data Set}

\bibitem[{DESI Collaboration}(2022)]{desilike}
{DESI Collaboration}. 2022, \texttt{desilike}, \url{https://github.com/cosmodesi/desilike}

\bibitem[{DESI Collaboration} {et~al.}(2025{\natexlab{a}})]{DESI_DR1}
{DESI Collaboration}, {Abdul-Karim}, M., {Adame}, A.~G., {et~al.} 2025{\natexlab{a}}, arXiv e-prints, arXiv:2503.14745

\bibitem[{DESI Collaboration} {et~al.}(2025{\natexlab{b}})]{DESI_DR2_I}
{DESI Collaboration}, {Abdul-Karim}, M., {Aguilar}, J., {et~al.} 2025{\natexlab{b}}, arXiv e-prints, arXiv:2503.14739

\bibitem[{DESI Collaboration} {et~al.}(2025{\natexlab{c}})]{DESI_DR2_CosmoConstrain}
{DESI Collaboration}, {Abdul-Karim}, M., {Aguilar}, J., {et~al.} 2025{\natexlab{c}}, arXiv e-prints, arXiv:2503.14738

\bibitem[{Dvali} {et~al.}(2000)]{DGP}
{Dvali}, G., {Gabadadze}, G., \& {Porrati}, M. 2000, Physics Letters B, 485, 208

\bibitem[{Fang} {et~al.}(2008)]{Fang+08}
{Fang}, W., {Wang}, S., {Hu}, W., {et~al.} 2008, \prd, 78, 103509

\bibitem[{Fang} {et~al.}(2020)]{Fang_2020}
{Fang}, X., {Krause}, E., {Eifler}, T., \& {MacCrann}, N. 2020, \jcap, 2020, 010

\bibitem[{Frieman} {et~al.}(2008)]{Fri+08}
{Frieman}, J.~A., {Turner}, M.~S., \& {Huterer}, D. 2008, \araa, 46, 385

\bibitem[{Frusciante} {et~al.}(2019)]{Frusciante+19}
{Frusciante}, N., {Peirone}, S., {Casas}, S., \& {Lima}, N.~A. 2019, \prd, 99, 063538

\bibitem[{Frusciante} \& {Perenon}(2020)]{EFTofDE_Review}
{Frusciante}, N., \& {Perenon}, L. 2020, \physrep, 857, 1

\bibitem[{Gil-Mar{\'\i}n} {et~al.}(2016)]{SDSS_FS_2015}
{Gil-Mar{\'\i}n}, H., {Percival}, W.~J., {Brownstein}, J.~R., {et~al.} 2016, \mnras, 460, 4188

\bibitem[{Gubitosi} {et~al.}(2013)]{Gubitosi_EFTofDE}
{Gubitosi}, G., {Piazza}, F., \& {Vernizzi}, F. 2013, \jcap, 2013, 032

\bibitem[{Hillebrandt} \& {Niemeyer}(2000)]{TyIaExpModels}
{Hillebrandt}, W., \& {Niemeyer}, J.~C. 2000, \araa, 38, 191

\bibitem[Horndeski(1974)]{Horndeski74}
Horndeski, G.~W. 1974, Int. J. Theor. Phys., 10, 363

\bibitem[{Hu} {et~al.}(2014{\natexlab{a}})]{EFTCAMB}
{Hu}, B., {Raveri}, M., {Frusciante}, N., \& {Silvestri}, A. 2014{\natexlab{a}}, \prd, 89, 103530

\bibitem[{Hu} {et~al.}(2014{\natexlab{b}})]{EFTCAMBNumericalNotes}
{Hu}, B., {Raveri}, M., {Frusciante}, N., \& {Silvestri}, A. 2014{\natexlab{b}}, {EFTCAMB/EFTCosmoMC: Numerical Notes v3.0}

\bibitem[{Hu} \& {Sawicki}(2007{\natexlab{a}})]{HuSaw07_new}
{Hu}, W., \& {Sawicki}, I. 2007{\natexlab{a}}, \prd, 76, 064004

\bibitem[{Hu} \& {Sawicki}(2007{\natexlab{b}})]{HuSaw07}
{Hu}, W., \& {Sawicki}, I. 2007{\natexlab{b}}, \prd, 76, 104043

\bibitem[{Huang}(2016)]{Huang16}
{Huang}, Z. 2016, \prd, 93, 043538

\bibitem[{Ishak} {et~al.}(2024)]{DESI_2024FSMG}
{Ishak}, M., {Pan}, J., {Calderon}, R., {et~al.} 2024, {Modified Gravity Constraints from the Full Shape Modeling of Clustering Measurements from DESI 2024}

\bibitem[{Joyce} {et~al.}(2015)]{Joy+15}
{Joyce}, A., {Jain}, B., {Khoury}, J., \& {Trodden}, M. 2015, \physrep, 568, 1

\bibitem[{Lai} {et~al.}(2025)]{DESI_2024FullModelingPyBird}
{Lai}, Y., {Howlett}, C., {Maus}, M., {et~al.} 2025, {A comparison between ShapeFit compression and Full-Modelling method with PyBird for DESI 2024 and beyond}

\bibitem[{Lesgourgues}(2011)]{class}
{Lesgourgues}, J. 2011, {The Cosmic Linear Anisotropy Solving System (CLASS) I: Overview}

\bibitem[Lewis \& Challinor(2016)]{CAMBWeb}
Lewis, A., \& Challinor, A. 2016, \texttt{CAMB}, \url{http://camb.info}

\bibitem[{Lewis} {et~al.}(2000)]{CAMB}
{Lewis}, A., {Challinor}, A., \& {Lasenby}, A. 2000, \apj, 538, 473

\bibitem[{Linder}(2003)]{Linder_2003}
{Linder}, E.~V. 2003, \prl, 90, 091301

\bibitem[{Linder}(2005)]{Linder05}
{Linder}, E.~V. 2005, \prd, 72, 043529

\bibitem[{Lombriser} {et~al.}(2009)]{Lombriser+09}
{Lombriser}, L., {Hu}, W., {Fang}, W., \& {Seljak}, U. 2009, \prd, 80, 063536

\bibitem[{Lu} {et~al.}(2025)]{lu2025preferenceevolvingdarkenergy}
{Lu}, Z., {Simon}, T., \& {Zhang}, P. 2025, {Preference for evolving dark energy in light of the galaxy bispectrum}

\bibitem[{Madhavacheril} {et~al.}(2024)]{Madhavacheril_2024}
{Madhavacheril}, M.~S., {Qu}, F.~J., {Sherwin}, B.~D., {et~al.} 2024, \apj, 962, 113

\bibitem[{Maus} {et~al.}(2025{\natexlab{a}})]{DESI_2024EFTComparison}
{Maus}, M., {Lai}, Y., {Noriega}, H.~E., {et~al.} 2025{\natexlab{a}}, {A comparison of effective field theory models of redshift space galaxy power spectra for DESI 2024 and future surveys}

\bibitem[{Maus} {et~al.}(2025{\natexlab{b}})]{DESI_2024AnaParamCompFullModelingVelocileptors}
{Maus}, M., {Chen}, S., {White}, M., {et~al.} 2025{\natexlab{b}}, {An analysis of parameter compression and Full-Modeling techniques with Velocileptors for DESI 2024 and beyond}

\bibitem[{Noriega} {et~al.}(2025)]{DESI_2024FullModelingFOLPS}
{Noriega}, H.~E., {Aviles}, A., {Gil-Mar{\'\i}n}, H., {et~al.} 2025, {Comparing Compressed and Full-Modeling analyses with FOLPS: implications for DESI 2024 and beyond}

\bibitem[{Perlmutter} {et~al.}(1999{\natexlab{a}})]{Perlmutter+99}
{Perlmutter}, S., {Aldering}, G., {Goldhaber}, G., {et~al.} 1999{\natexlab{a}}, \apj, 517, 565

\bibitem[{Perlmutter} {et~al.}(1999{\natexlab{b}})]{Perlmutter_1999}
{Perlmutter}, S., {Aldering}, G., {Goldhaber}, G., {et~al.} 1999{\natexlab{b}}, \apj, 517, 565

\bibitem[{Phillips} {et~al.}(1999)]{Phillips_1999}
{Phillips}, M.~M., {Lira}, P., {Suntzeff}, N.~B., {et~al.} 1999, \aj, 118, 1766

\bibitem[{Planck Collaboration} {et~al.}(2016)]{planck2015}
{Planck Collaboration}, {Ade}, P.~A.~R., {Aghanim}, N., {et~al.} 2016, \aap, 594, A14

\bibitem[{Planck Collaboration} {et~al.}(2020)]{planck2018}
{Planck Collaboration}, {Aghanim}, N., {Akrami}, Y., {et~al.} 2020, \aap, 641, A6

\bibitem[{Qu} {et~al.}(2024)]{Qu_2024}
{Qu}, F.~J., {Sherwin}, B.~D., {Madhavacheril}, M.~S., {et~al.} 2024, \apj, 962, 112

\bibitem[{Ramirez-Solano} {et~al.}(2025)]{DESI_2024FullModelingParamComp}
{Ramirez-Solano}, S., {Icaza-Lizaola}, M., {Noriega}, H.~E., {et~al.} 2025, {Full Modeling and parameter compression methods in configuration space for DESI 2024 and beyond}

\bibitem[{Raveri} {et~al.}(2014{\natexlab{a}})]{Raveri+14}
{Raveri}, M., {Hu}, B., {Frusciante}, N., \& {Silvestri}, A. 2014{\natexlab{a}}, \prd, 90, 043513

\bibitem[{Raveri} {et~al.}(2014{\natexlab{b}})]{EFTCAMB_ConDE}
{Raveri}, M., {Hu}, B., {Frusciante}, N., \& {Silvestri}, A. 2014{\natexlab{b}}, \prd, 90, 043513

\bibitem[{Riess} {et~al.}(1998{\natexlab{a}})]{Reiss+98}
{Riess}, A.~G., {Filippenko}, A.~V., {Challis}, P., {et~al.} 1998{\natexlab{a}}, \aj, 116, 1009

\bibitem[{Riess} {et~al.}(1998{\natexlab{b}})]{Riess_1998}
{Riess}, A.~G., {Filippenko}, A.~V., {Challis}, P., {et~al.} 1998{\natexlab{b}}, \aj, 116, 1009

\bibitem[{Riess} {et~al.}(2011)]{Riess_2011}
{Riess}, A.~G., {Macri}, L., {Casertano}, S., {et~al.} 2011, \apj, 730, 119

\bibitem[{Sawicki} {et~al.}(2013)]{Sawicki_2013}
{Sawicki}, I., {Saltas}, I.~D., {Amendola}, L., \& {Kunz}, M. 2013, \jcap, 2013, 004

\bibitem[{Taule} {et~al.}(2025)]{SDSS_FS_MG}
{Taule}, P., {Marinucci}, M., {Biselli}, G., {Pietroni}, M., \& {Vernizzi}, F. 2025, \jcap, 2025, 036

\bibitem[{Weinberg}(1989)]{Weinberg89}
{Weinberg}, S. 1989, Reviews of Modern Physics, 61, 1

\bibitem[Wilks(1938)]{WilksTheorem}
Wilks, S.~S. 1938, The Annals of Mathematical Statistics, 9, 60

\bibitem[{Will}(2014)]{Will14}
{Will}, C.~M. 2014, Living Reviews in Relativity, 17, 4

\bibitem[{Will}(2018)]{Will18}
{Will}, C.~M. 2018, {Theory and Experiment in Gravitational Physics}

\bibitem[{Zhao} {et~al.}(2009{\natexlab{a}})]{Zhao+09}
{Zhao}, G.-B., {Pogosian}, L., {Silvestri}, A., \& {Zylberberg}, J. 2009{\natexlab{a}}, \prd, 79, 083513

\bibitem[{Zhao} {et~al.}(2009{\natexlab{b}})]{MGCAMB}
{Zhao}, G.-B., {Pogosian}, L., {Silvestri}, A., \& {Zylberberg}, J. 2009{\natexlab{b}}, \prd, 79, 083513

\bibitem[{Zlatev} {et~al.}(1999)]{Zlatev99}
{Zlatev}, I., {Wang}, L., \& {Steinhardt}, P.~J. 1999, \prl, 82, 896

\bibitem[{Zumalac{\'a}rregui} {et~al.}(2017)]{hiclass}
{Zumalac{\'a}rregui}, M., {Bellini}, E., {Sawicki}, I., {Lesgourgues}, J., \& {Ferreira}, P.~G. 2017, \jcap, 2017, 019

\end{thebibliography}

\end{document}